\documentclass[10pt,conference]{IEEEtran}
\usepackage{latexsym,amssymb}
\title{Codes for error detection, good or not good}
\author{\authorblockN{Irina Gancheva}
\authorblockA{Department of Informatics\\ University of Bergen, Norway \\
Email: Irina.Gancheva@ii.uib.no}
\and
\authorblockN{Torleiv Kl{\o}ve}
\authorblockA{Department of Informatics\\ University of Bergen, Norway\\
Email: Torleiv.Klove@ii.uib.no} 
 }
\newtheorem{theorem}{Theorem}
\newtheorem{lemma}{Lemma}
\newtheorem{corollary}{Corollary}
\newtheorem{example}{Example}
\def\slutt {\hfill $\Box$\bigskip }

\begin{document}
\maketitle

\begin{abstract}
Linear codes for error detection on a $q$-ary symmetric channel are studied.
It is shown that for given dimension $k$ and minimum distance $d$, there exists a value $\mu(d,k)$ such that if $C$ is a code of length $n\ge \mu(d,k)$, then neither 
$C$ nor its dual $C^\perp$ are good for error detection. For $d\gg k$ or $k\gg d$ good approximations for $\mu(d,k)$ are given. A generalization to non-linear codes is also given.
\end{abstract}

\section{Introduction}
The $q$-ary symmetric channel with symbol probability $p$, where 
$0\le p \le \frac{q-1}{q}$, is defined as follows:
symbols from $GF(q)$ are transmitted over the channel, and
\[P(b \mbox{ received}\mid a \mbox{ sent})=\left\{
\begin{array}{ll}
1-p & b=a\\
\frac{p}{q-1} & b\ne a
\end{array}\right.\]

Suppose that a $q$-ary linear $[n,k,d]$ code $C$ is used for error detection for 
transmission over the $q$-ary symmetric channel with symbol error probability $p$. 
The probability of undetected error for $C$ is denoted by $P_{\rm ue}(C,p)$.
For most codes we are not able to determine the value of $P_{\rm ue}(C,p)$ exactly.
Therefore, it is useful to have estimates. $C$ is called \emph{good}
(for error detection) if
\begin{equation}
\label{good}
P_{\rm ue}(C,p)\le P_{\rm ue}(C,(q-1)/q)=\frac{q^k-1}{q^n}.
\end{equation}
Only some codes have this property. The purpose of this paper is to show that, for given dimension $k$ and minimum distance $d$, there exists a value $\mu(d,k)$ such that if $n\ge \mu(d,k)$ and $C$ is an $[n,k,d]$ code, then neither 
$C$ nor its dual $C^\perp$ are good for error detection. Further, for $d\gg k$ or $k\gg d$ we give good approximations of $\mu(d,k)$. At the end of the paper we give similar results for non-linear codes.

\section{Existence of $\mu(d,k)$}
 If $A_i$, $i=0,1,\ldots, n$ is the weight distribution of $C$, then
the probability of undetected error for $C$ and $C^\perp$ on a symmetric channel with 
symbol error probability $p$ is given by
\begin{equation}
\label{pueC}
P_{\rm ue}(C,p)=\sum_{i=1}^n A_i \Bigl(\frac{p}{q-1}\Bigr)^i(1-p)^{n-i},
\end{equation}
\begin{equation}
\label{pueCdual}
P_{\rm ue}(C^\perp,p)=q^{-k}\sum_{i=0}^n A_i (1-Qp)^i-(1-p)^n,
\end{equation}
where $Q=q/(q-1)$.

Note that $C$ is good if
$P_{\rm ue}(C,p)\leq q^{k-n}-q^{-n}$ for all $p\in (0,(q-1)/q)$. We call $C$ \emph{bad} if
$P_{\rm ue}(C,p)\ge q^{k-n}$ for some $p\in (0,(q-1)/q)$. Note that being bad is a stronger condition than not being good. We also note that most codes are either good or bad, but a code may be neither.

\begin{lemma}
\label{bad}
$C$ is bad if and only if $C^\perp$ is bad.
\end{lemma}
Proof. Suppose $C$ is bad. By definition there is a $p\in (0,(q-1)/q)$ such that 
$P_{\rm ue}(C,p)\ge q^{k-n}$. Define $\pi$ by
\[1-Q\pi=\frac{p}{(q-1)(1-p)},\]
that is
\[\pi=\frac{(q-1)-qp}{q(1-p)}\quad\mbox{and}\quad 1-\pi=\frac{1}{q(1-p)}.\]
Then
\begin{eqnarray*}
q^{k-n} &\leq &  P_{\rm ue}(C,p)\\
  &=& (1-p)^n \sum_{i=1}^n A_i \Bigl(\frac{p}{(q-1)(1-p)}\Bigr)^i \\
  &=& (1-p)^n \Bigl(\sum_{i=0}^n A_i (1-Q\pi)^i -1\Bigr)\\
  &=& (1-p)^n \Bigl(q^{k} P_{\rm ue}(C^\perp,\pi) + (1-\pi)^n -1 \Bigr) \\
  &=& (1-p)^n \Bigl(q^{k} P_{\rm ue}(C^\perp,\pi) -1 \Bigr) \\
  && \vspace{2 cm} +q^{k}(1-\pi)^n(1-p)^n\\
  &=& (1-p)^n \Bigl(q^{k} P_{\rm ue}(C^\perp,\pi) -1 \Bigr) +q^{k-n}
\end{eqnarray*}
and hence $P_{\rm ue}(C^\perp,\pi)\ge q^{-k}$, that is, $C^\perp$ is bad.
Since $C^{\perp\perp}=C$ we get the if and only if.\slutt

Remark 1. For $q=2$, Lemma \ref{bad} is Theorem 3.4.2 part 1 in \cite{kk}.
The proof for general $q$ given above is a straightforward generalization of the
proof for $q=2$ given in \cite{kk}. 

Remark 2. It is not the case that $C$ not good implies that $C^\perp$ is not good.
\bigskip

We want to find sufficient conditions for a code to be bad. If the minimum distance of 
$C$ is $d$ (that is, $A_i=0$ for $i=1,2,\ldots ,d-1$ and $A_d>0$), then $A_d\ge q-1$ by the linearity and hence
\[P_{\rm ue}(C^\perp,p)\ge q^{-k}+q^{-k}(q-1) (1-Qp)^d-(1-p)^n \ge q^{-k}\]
if $(q-1)(1-Qp)^d\ge q^{k}(1-p)^n$. Taking logarithms, the last condition is equivalent
to 
\[\ln(q-1)+d \ln(1-Qp)\ge n\ln(1-p)+k\ln(q).\]
 Hence, combining with Lemma \ref{bad},
 we get the following lemma.
\begin{lemma}
\label{nglemma}
If 
\[
n\ge h(p)= \frac{d\ln(1-Qp)-k\ln(q)+\ln(q-1)}{\ln(1-p)},
\]
then both $C$ and $C^\perp$ are bad.
\end{lemma}

We will use the notations $\kappa=k\ln(q)-\ln(q-1)$, 
\[f(p)=\frac{\ln(1-Qp)}{\ln(1-p)},\mbox{ and }g(p)=\frac{-1}{\ln(1-p)}.\]
Then
\begin{equation}
\label{ngc}
h(p)= d\, f(p)+\kappa\, g(p).
\end{equation}

Any choice of $p$, $0<p<(q-1)/q$ now gives a proof of the existence of $\mu(k,d)$ such that if $n\ge \mu(d,k)$ and $C$ is an $[n,k,d]$ code, then neither 
$C$ nor its dual $C^\perp$ are good for error detection.

To get the strongest result from the lemma, we want to find the $p$ that minimizes
$h(p)$. 
The function $f(p)$ is increasing on $(0,(q-1)/q)$, it approaches 
the value Q when $p\rightarrow 0+$, and it approaches infinity when $p\rightarrow (q-1)/q-$.
Moreover,
\[f'(p)=\frac{-Q(1-p)\ln(1-p)+(1-Qp)\ln(1-Qp)}{(1-p)(1-Qp)\ln(1-p)^2},\]
and
\[f''(p)=\frac{f_1(p)}{-(1-p)^2 (1-Qp)^2 (\ln(1-p))^3 },\]
where
\begin{eqnarray*}
f_1(p) &=& Q^2 (1-p)^2(\ln(1-p))^2 \\
       && +2Q(1-p)(1-Qp)\ln(1-p) \\
       && -2(1-Qp)^2\ln(1-Qp) \\
       && -(1-Qp)^2\ln(1-p)\ln(1-Qp)\\
       &>& 0 
\end{eqnarray*}
for all $p\in (0,(q-1)/q)$. Hence $f$ is convex on $(0,(q-1)/q)$.
Similarly, the function $g(p)$ is decreasing on $(0,(q-1)/q)$, it approaches 
infinity when $p\rightarrow 0+$, and it takes the value $-1/\ln(q)$ for $p= (q-1)/q$. Moreover,
\[g'(p)=\frac{1}{(1-p)\ln(1-p)^2}=\frac{1-Qp}{(1-p)(1-Qp)\ln(1-p)^2},\]
\[g''(p)=\frac{-(2+\ln(1-p))}{(1-p)^2 (\ln(1-p))^3}>0\]
 for all $p\in (0,(q-1)/q)$, and so $g(p)$ is also convex on $(0,(q-1)/q)$.
This implies that the combined function $h(p)$ is also convex on $(0,(q-1)/q)$
since $\kappa>0$,
and it takes its minimum somewhere in $(0,(q-1)/q)$.
We denote this minimum by $\mu(d,k)$. From Lemmas \ref{bad} and \ref{nglemma} we
get the following corollary. 
\begin{corollary}
If $n\ge \mu(d,k)$, then neither $C$ nor $C^\perp$ are good for error detection.
\end{corollary}
We next find approximations for $\mu(d,k)$ when $d\gg k$ or $k\gg d$.
We denote by $p_m$ the value of $p$ where $h(p)$ has its minimum 
(the minimum is by definition $\mu(d,k)$).

\section{Approximations for $\mu(d,k)$ when $d\gg k$}

\begin{theorem}
\label{t1}
Assume that $d\rightarrow \infty$ and $k/d\rightarrow 0$. Let 
\[\kappa=k\ln(q)-\ln(q-1) \mbox{ and } y=\sqrt{\frac{\kappa}{2dQ(Q-1)}}.\]
There exist numbers
$a_i$ and $b_i$ for $i=1,2,\ldots$ such that, 
for any $r\ge 0$,
\[p_m=\sum_{i=1}^r a_i y^i +O(y^{r+1}),\]
and 
\[\mu(d,k)=dQ+2dQ(Q-1)\sum_{i=1}^r b_i y^i+O(y^{r+1}).\]
The first few $a_i$ and $b_i$ are given by the following table:
\begin{eqnarray*}
a_1 &=& 2,\\
a_2 &=& -(8Q+2)/3,\\
a_3 &=& (26Q^2+22Q-1)/9,\\
a_4 &=& -(368Q^3+708Q^2-12Q+8)/135,\\
b_1 &=& 1,\\
b_2 &=& (2Q-1)/3,\\
b_3 &=& (2Q^2-2Q-1)/18,\\
b_4 &=& -(4Q^3-6Q^2-6Q+4)/135.
\end{eqnarray*}
\end{theorem}
Proof: First we note that $\kappa=2dQ(Q-1)y^2$ and so
\[h(p)=d\, \frac{\ln(1-Qp)-2Q(Q-1)y^2}{\ln(1-p)},\]
and
\[h'(p)=d\, \frac{H(p,y)}{(1-p)(1-Qp)(\ln(1-p))^2},\]
where
\begin{eqnarray*}
H(p,y) &=& -Q(1-p)\ln(1-p)+(1-Qp)\ln(1-Qp) \\
     && -2Q(Q-1)y^2 (1-Qp).
\end{eqnarray*}
Hence $h'(p)=0$ if $H(p,y)=0$.
Taking the Taylor expansion of $H(\sum a_i y^i,y)$ we get
\begin{eqnarray*}
H(\sum a_i y^i,y) &=& \frac{a_1^2 - 4}{4}y^2 \\
                  && +\frac{a_1}{6} (Qa_1^2+a_1^2+6 a_2+12Q) y^3 + \cdots
\end{eqnarray*}
All coefficients should be zero. In particular, the coefficient of $y^2$ shows that $a_1^2=4$. Since $a_1 y^2$ is the dominating term in the expression for $p$ when $y$ is small and $p>0$, we must have $a_1>0$ and so $a_1=2$.
Next the coefficient of $y^3$ shows that 
$a_2=-(16Q+4)/6$. In general,
 we get equations in the $a_i$ which can be used to determine the $a_i$ recursively. The recursions seems to be quite complicated in general and we have not found an explicit general expression for $a_i$.
Substituting the expression for $p$ into $h(p)$ and taking Taylor expansion, we get the expression for $\mu(d,k)$.
\slutt

Remark. We do not know when the infinite series $\sum_{i=1}^\infty a_i y^i$ and 
$\sum_{i=1}^\infty b_i y^i$ converge (this may depend on $q$ and $p$).

\begin{example}
Consider $q=2$, $d=1000$ and $k=2$. Solving $h'(p)=0$ numerically, we get 
$p\approx 0.0352540$ and $\mu(1000,2)\approx 2075.8565430$.
Taking one, two, three, and four terms respectively in the expression for $\mu(1000,2)$ in Theorem \ref{t1},
we get the following approximations:
\[\begin{array}{c|l}
\mbox{no. of terms} & \mbox{value} \\ \hline
1 & 2000 \\
2 & 2074.4659482 \\
3 & 2075.8522426 \\
4 & 2075.8565439
\end{array}\]
\end{example}

\section{Approximations for $\mu(d,k)$ when $k\gg d$}

\begin{theorem}
\label{t2}
Assume that $k\rightarrow \infty$ and $d/k\rightarrow 0$. Let
\[\begin{array}{ll}
\kappa=k\ln(q)-\ln(q-1),  & \lambda=\ln(q), \\
\theta=d/\kappa,  & \Lambda=\ln(\theta \lambda/(q-1))
\end{array}\]
There exist polynomials $A_i(x)$ and $B_i(x)$ 
for $i=0,1,2,\ldots$ such that, 
for any $r\ge 0$,
\[1-Qp_m=\frac{\lambda}{q-1}\sum_{i=1}^r A_i(\Lambda) \theta^i 
 +O(\theta^{r+1}),\]
and 
\[\mu(d,k)=\frac{\kappa}{\lambda}\sum_{i=0}^r B_i(\Lambda) \theta^i+O(\theta^{r+1}).\]
The first few $A_i(x)$ and $B_i(x)$ are given by the following table:
\begin{eqnarray*}
A_1(x) &=& 1,\\
A_2(x) &=& x+\lambda-1,\\
A_3(x) &=& (2x^2+(4\lambda-2)x+2\lambda^2-3\lambda)/2,\\
B_0(x) &=& 1,\\
B_1(x) &=& -(x-1),\\
B_2(x) &=& -(2x-\lambda+2)/2,\\
B_3(x) &=& -(3x^2+3\lambda x +\lambda^2 -3)/6.
\end{eqnarray*}
\end{theorem}
Proof: Let $\eta=q-1$ and $\pi=1-Qp$. Then $1-p=(1+\eta\pi)/q$ and
$Q(1-p)=(1+\eta\pi)/\eta$.
Hence
\[h'(p)=\kappa\, \frac{G(\pi,\theta)}{(1-p)(1-Qp)\ln(1-p)^2},\]
where
\[G(\pi,\theta)=-\theta\frac{1+\eta\pi}{\eta}
\ln\Bigl(\frac{1+\eta\pi}{q}\Bigr)+\theta \pi \ln(\pi)-\pi.\]
Therefore, $h'(p)=0$ if and only if $G(\pi,\theta)=0$. 

If $\pi \rightarrow 0+$, then $\ln((1+\eta\pi)/q)\rightarrow -\ln(q)$ and $\pi \ln(\pi)\rightarrow 0$.
Hence, for small $\pi$,
\[0=\frac{G(\pi,\theta)}{\pi}\approx \frac{\ln(q)}{\eta}-\frac{\pi}{\theta}.\]
Therefore,  $\pi\approx \theta\lambda/\eta$. We write 
$\pi=\theta\lambda(1+y)/\eta$ 
(where $y$ will depend on $\theta$). This implies that 
\[\ln(\pi)=\Lambda+\ln(1+y).\]
 Hence, if $g(y)=G(\pi,\theta)$ we get
\begin{eqnarray*}
g(y) &=& \frac{\theta}{\eta}\Bigl\{- (1+\theta\lambda(1+y))\ln(1+\theta\lambda(1+y)) \\
     && \qquad +(1+\theta\lambda(1+y))\lambda \\
     && \qquad +\theta \lambda (1+y)(\Lambda+\ln(1+y)) \\
     && \qquad -\lambda (1+y)\Bigr\}.
\end{eqnarray*}
We now write $y=\sum_{i=1}^r \alpha_i(\Lambda) \theta^i +O(\theta^{r+1})$.
Formally treating $\Lambda$ as if it were a constant, we can take the Taylor expansion of $g(y)$ in 
terms of $\theta$ and we get an expansion of the form 
$\sum_{i=1}^\infty c_i \theta^i$, where the $c_i$ are polynomials of $\Lambda$. 
Since these polynomials $c_i$ must be identically zero, we get equations to determine the polynomials $A_i(x)$. 
Substituting the series of $p_m$ into $h(p)$ and taking the Taylor expansion, we get the series of $\mu(d,k)$.\slutt

\noindent Remark. To formally justify that we treat $\Lambda$ as if it were a constant, we should prove that $\Lambda$ is algebraically independent of the other quantities involved, that is, there is no non-trivial polynomial equation in $\Lambda$ with coefficients expressed as rational functions of the remaining 
quantities. We have not done this, but it highly likely that it is true.

As to convergence, the situation is similar to Theorem \ref{t1}.

\begin{example}
Consider $q=2$, $d=2$ and $k=1000$. Solving $h'(p)=0$ numerically, we get 
$p\approx 0.4990185$ and $\mu(2,1000)\approx 1020.8737393$.
Taking one, two, three, and four terms respectively in the expression for $\mu(2,1000)$ in Theorem \ref{t2},
we get the following approximations:
\[\begin{array}{c|l}
\mbox{no. of terms} & \mbox{value} \\ \hline
1 & 1000 \\
2 & 1020.8169587 \\
3 & 1020.8741383 \\
4 & 1020.8737362
\end{array}\]
\end{example}

\section{Non-linear codes}

To get results for a non-linear code $C$, we have to do some modifications.
In stead of the weight distribution, we let $\{A_i\}$ denote the distance distribution (for linear codes these are the same). Then equation (\ref{pueC}) is still valid for non-linear codes. 

Non-linear codes do not have dual codes so results like 
(\ref{pueCdual}) and Lemma \ref{bad} do not make sense in the non-linear case. However, if we let $k=\log_q(|C|)$ and define  
\[P^\perp_{\rm ue}(C,p)=q^{-k}\sum_{i=0}^n A_i (1-Qp)^i-(1-p)^n,\]
this is a welldefined function, and for linear codes 
$P_{\rm ue}(C^\perp,p)=P^\perp_{\rm ue}(C,p)$. Moreover, using
$P^\perp_{\rm ue}(C,\pi)$ in stead of $P_{\rm ue}(C^\perp,\pi)$
and $P^\perp_{\rm ue}(C,p)$ in stead of $P_{\rm ue}(C^\perp,p)$
in the proof of Lemmas \ref{bad} and \ref{nglemma}, we get the following result.

\begin{lemma}
\label{baddual}
If $C$ is a bad code of length $n$ and size $M$, 
and $1-Q\pi=\frac{p}{(q-1)(1-p)}$, then
\[P^\perp_{\rm ue}(C,\pi)\ge 1/M.\]
\end{lemma}

From this we get a lemma similar to Lemma \ref{nglemma}.
However, whereas for linear codes we have $A_d\ge q-1$, 
for non-linear codes we only know in general that $A_d\ge 1/M$.
Hence we get the following weaker version of Lemma \ref{nglemma}
in general.

\begin{lemma}
\label{nnglemma}
If $C$ is a code of length $n$ and size $M$, and
\begin{equation}
\label{nngc}
n\ge h_N(p)= \frac{d\ln(1-Qp)-2\log_q(M)\ln(q)}{\ln(1-p)},
\end{equation}
then $C$ is bad.
\end{lemma}

The analysis of the cases $d\gg k$ and $k\gg d$ to find the best choice 
of $p$ is similar to the linear case, the only difference is that for the general case $\kappa=2k\ln(q)$. With this modification, Theorems \ref{t1} and \ref{t2} are valid also for non-linear codes. Let $\mu_N(d,k)$ denote the minimum
of $h_N(p)$ (for $\kappa=2k\ln(q)$) and let $p_N$ be the value of $p$ that gives this minimum. 

\begin{theorem}
\label{t1n}
Assume that $d\rightarrow \infty$ and $k/d\rightarrow 0$. Let 
\[\kappa=2k\ln(q) \mbox{ and } y=\sqrt{\frac{\kappa}{2dQ(Q-1)}}.\]
For any $r\ge 0$,
\[p_N=\sum_{i=1}^r a_i y^i +O(y^{r+1}),\]
and 
\[\mu_N(d,k)=dQ+2dQ(Q-1)\sum_{i=1}^r b_i y^i+O(y^{r+1}).\]
where $a_i$ and $b_i$ are the numbers given in Theorem \ref{t1}.
\end{theorem}
\newpage

\begin{theorem}
\label{t2n}
Assume that $k\rightarrow \infty$ and $d/k\rightarrow 0$. Let
\[\begin{array}{ll}
\kappa=2k\ln(q),  & \lambda=\ln(q), \\
\theta=d/\kappa,  & \Lambda=\ln(\theta \lambda/(q-1))
\end{array}\] 
For any $r\ge 0$,
\[1-Qp_N=\frac{\lambda}{q-1}\sum_{i=1}^r A_i(\Lambda) \theta^i 
 +O(\theta^{r+1}),\]
and 
\[\mu_N(d,k)=\frac{\kappa}{\lambda}\sum_{i=0}^r B_i(\Lambda) \theta^i+O(\theta^{r+1}),\]
where 
$A_i(x)$ and $B_i(x)$ are the polynomials given in Theorem \ref{t2}.
\end{theorem}

\begin{example}
For the values of $d$ and $k$ (and $q=2$) considered in the examples for linear codes, we get the following approximate values for the bounds:
\[\begin{array}{rr|ll}
   d &    k & \mu(d,k) & \mu_N(d,k) \\ \hline
1000 &    2 & 2075.86 & 2108.10 \\
   2 & 1000 & 1020.87 & 2022.85 
\end{array}\]
This illustrate the results we have found that for large $d$, $\mu(d,k)$ and $\mu(d,k)_N$ are both approximately $2d$ (taking only the first term of the approximation)
and for large $k$, $\mu(d,k)$ is approximately $k$ whereas $\mu(d,k)_N$ is approximately $2k$ (reflecting the fact that $\kappa$ is approximately twice as large in the non-linear case).

\end{example}

\section*{Acknowledgment}
The research was supported by The Norwegian Research Council.

\end{document}